\documentclass[twocolumn,english,showpacs]{revtex4}
\usepackage{amsmath,amssymb}
\usepackage[dvips]{graphics}
\def\imo{i}
\begin{document}

\title{Stability of multidimensional black holes: complete numerical analysis}
\author{R. A. Konoplya}
\email{konoplya@fma.if.usp.br}
\author{A. Zhidenko}
\email{zhidenko@fma.if.usp.br}
\affiliation{Instituto de F\'{\i}sica, Universidade de S\~{a}o Paulo \\
C.P. 66318, 05315-970, S\~{a}o Paulo-SP, Brazil}

\pacs{04.30.Nk,04.50.+h}

\begin{abstract}
We analyze evolution of gravitational perturbations of
$D$-dimensional Schwarzschild, Reissner-Nordstr\"om, and
Reissner-Nordstr\"om-de Sitter black holes. It is known that the
effective potential for the scalar type of gravitational
perturbations has negative gap near the event horizon. This gap,
for some values of the parameters $Q$ (charge), $\Lambda$
(cosmological constant) and $D$ (number of space-time dimensions),
cannot be removed by S-deformations. Thereby, there is no proof of
(in)stability for those cases. In the present paper, by an
extensive search of quasinormal modes, both in time and frequency
domains, we shall show that spherically symmetric static black
holes with arbitrary charge and positive (de Sitter) lambda-term
are stable for $D=5, 6, \ldots11$. In addition, we give a
complete numerical data for all three types (scalar, vector and
tensor) of gravitational perturbations for multi-dimensional black
holes with charge and $\Lambda$-term. The influence of charge,
$\Lambda$-term and number of extra dimensions on black hole
quasinormal spectrum is discussed.
\end{abstract}

\maketitle

\section{Introduction}

With the development of the string theory and higher-dimensional
brane-world models, black holes in more than four space-time
dimensions became a key subject in the modern high energy physics.
Nevertheless, one of the first $D$-dimensional generalizations of
the Schwarzschild metric was done, yet, in 1963 by Tangherlini
\cite{Tangherlini:1963bw}.
This solution has attracted considerable interest recent years, in
the context of brane-world models \cite{brane-world}, and, for
asymptotically anti-de Sitter case, within AdS/CFT correspondence
\cite{AdSCFT}. Different generalizations of the Tangherlini metric were found
\cite{generalD}.
and their properties were investigated in  \cite{propertyD}. One
of the most important property of a black hole metric is
stability. Space-times unstable against small metric perturbations
should not exist, or need some other, probably quantum, mechanisms
to provide stability. If a black hole is unstable, small
perturbations must grow with time, being governed by
characteristic complex frequencies $\omega$, called {\it
quasinormal modes}. These frequencies form a spectrum of the
response of a black hole to the external perturbation.
Contribution of all modes can be seen in the {\it time domain}
\cite{Price-Pullin}. Until now, during recent few years, evolution of
perturbations of the $D$-dimensional black holes was investigated in
a number of works. A lot of papers were devoted to perturbations
of scalar test fields propagating in the vicinity of spherically
symmetric black holes \cite{scalarD}. Scalar QNMs of
higher-dimensional Kerr black holes were analyzed in
\cite{scalarD-Kerr}. Quasinormal spectrum of Standard Model fields for brane-localized
black holes were investigated in
\cite{Berti:2003yr,Abdalla:2006qj,Kanti:2005xa,Kanti:2006ua}.
In \cite{Kanti:2005xa}, a variety of brane-localized
backgrounds (Schwarzschild, Reissner-Nordstr\"om and
Schwarzschild-(anti) de Sitter) were investigated, and it was
shown that an increase in the number of hidden, extra dimensions
results in the faster damping of all fields living on the brane.
In \cite{Kanti:2006ua} rotating black holes on the brane were
considered, and, it was shown that the brane-localized field
perturbations are longer-lived when the higher-dimensional black
hole rotates faster.

In the context of Tev-gravity scenarios, the gravitational energy
loss in high energy particles collisions was considered in
\cite{Berti:2003si}, where was found energy spectra, total energy
and angular distribution of the emitted gravitational radiation.
The further steps were analysis of the graviton emission form the
higher dimensional black holes \cite{Cornell:2005ux} and the
close limit analysis for head-on collision of two black holes in
higher dimensions \cite{Yoshino:2005ps,Yoshino:2006kc}.

In addition, some analytical solutions were found in different
$D>4$ space-times \cite{D-exact}. The high overtone asymptotics
for perturbations of $D$-dimensional black holes were considered in
\cite{highND}.

After appearing of the paper of Kodama and Ishibashi
\cite{Kodama:2003kk,Kodama:2003jz,Ishibashi:2003ap}, where gravitational perturbations of the
$D$-dimensional spherically symmetric black hole with charge and
$\lambda$-term were reduced to the Schrodinger wave-like
equations:
\begin{equation}\label{wavelike}
\frac{d^{2} \Psi_{i}}{d r_{*}^{2}} + (\omega^{2} - V_{i}(r))\Psi_{i}=0,
\end{equation}
the low-laying gravitational quasinormal modes were found found in
\cite{Konoplya:2003dd} with the help of the WKB method \cite{WKB}. In (\ref{wavelike}), $i$ runs
$T$, $V$, $S$, because for $D>4$, in addition to the well-known in
four dimensions axial and polar perturbations, which are vector
and scalar types of gravitational perturbations respectively,
there appears a new type of gravitational perturbations called
tensor one. For $D$-dimensional Reissner-Nordstr\"om-(anti)-de Sitter
black holes the tensor type of gravitational perturbations
coincide with perturbations of test scalar filed. In the Einstein
theory with the Gauss-Bonnet term, from works \cite{Dotti:2005sq,Konoplya:2005sy,Gleiser:2005ra},
it follows that this coincidence does not
take place.

An important feature of the gravitational perturbations is that
the effective potentials which governs the scalar type of the
perturbations has negative gap for $D>7$, and this gap becomes
deeper when $D$ is increased. If the effective potential
$V_{i}(r)$ is positive definite everywhere outside the black hole,
event horizon, the differential operator
$$\frac{d^{2}}{dr_{*}^{2}} + \omega^{2}$$
is positive self-adjoint operator in the Hilbert space of the
square integrable functions of $r^{*}$. As a result, any solution
of the wave equation with compact support is bounded, what implies
stability. In some cases this negative gap of the potential can be
removed by the so-called S-deformation of the potential which does
not change the property of stability \cite{Ishibashi:2003ap}. In
this way Ishibashi and Kodama in \cite{Kodama:2003kk,Kodama:2003jz,Ishibashi:2003ap}, showed that
$D$-dimensional Schwarzschild black holes are stable against all
three types of gravitational perturbations. Yet, the potential for
scalar type of gravitational perturbations of Reissner-Nordstr\"om
black holes cannot be made positive definite by S-deformation, and
therefore {\it the question of stability of charged
multidimensional black holes left open}. That is why to judge
about (in)stability of black holes we have to consider evolution
of gravitational perturbations of $D$-dimensional black hole with
special emphasis in scalar type of gravitational perturbations.

Note, that with the help of the WKB method, one can find only low
laying modes for not very large $D$ \cite{Konoplya:2003dd}.
Therefore we have to apply here the Frobenius method
\cite{Leaver:1985ax}. Yet, even Frobenius method shows rather bad
convergence for higher $D$ and, this made us to apply,
alternatively, the time-domain approach. In the present paper we
have found all three types of gravitational quasinormal modes and
have shown that {\it the D=5-11 Reissner-Nordstr\"om(-de Sitter)
black holes are stable for any values of charge or
$\Lambda$-term}.

The paper is organized as follows. In Sec. \ref{formulae::sec} we give the basic
formulas and discuss the range of parameters for the $D$-dimensional
Reissner-Nordstr\"om(-de Sitter) metric and of its gravitational
perturbations. Sec \ref{numericalmethods::sec} describes the numerical method we used in
time and frequency domain. Sec \ref{numericalmethods::sec} relates the obtained
quasinormal modes and time-domain profiles for
Reissner-Nordstr\"om(-de Sitter) black holes in $D=5, 6, \ldots11$
space-time dimensions. Finally we discuss the obtained results and
conclude that all considered cases are stable.


\section{Basic formulas}\label{formulae::sec}

The metric of the $D$-dimensional Reissner-Nordstr\"om-de-Sitter black
holes is given by the formula

\begin{equation}\label{metric}
ds^2=f(r)dt^2-\frac{dr^2}{f(r)}-r^2(d\theta^2+sin^2\theta
d\phi^2).
\end{equation}
where
\begin{equation}\label{metric-function}
f(r)=1-X+Z-Y,
\end{equation}
$$X=\frac{2M}{r^{d-1}},\qquad Y=\frac{2\Lambda r^2}{d(d+1)}, \qquad Z=\frac{Q^2}{r^{2d-2}},$$

The wave equation can be re-written in the form
\begin{equation}\label{hyperbolic}
\left(\frac{\partial^2}{\partial t^2}-\frac{\partial^2}{\partial r_*^2}\right)\Psi(t,r)=-V(r)\Psi(t,r),
\end{equation}
where \emph{the tortoise coordinate} $r_*$ is defined as
\begin{equation}\label{tortoise}
dr_*=\frac{dr}{f(r)}.
\end{equation}

The effective potentials for tensor
\begin{equation}
\label{potential-tensortype}
V_T(r)=\frac{f(r)}{r^2}\left(\lambda+d+\frac{f'(r)rd}{2}+\frac{f(r)d(d-2)}{4}\right)
\end{equation}
 and vector gravitational perturbations
\begin{eqnarray}
\nonumber
V_V(r)=\frac{f(r)}{r^2}\left(\lambda+d\pm\sqrt{X^2\frac{(d^2-1)^2}{4}+2Z\lambda d(d-1)}\right.&&\\
\label{potential-vectortype} +\left.\frac{d(d-2)(1-Y)+Zd(5d-2)-X(d^2+2)}{4}\right)&&
\end{eqnarray}
are designated as $V_{T}$, $V_{V}$ respectively.

The effective potential for the scalar type is given by
\begin{equation}
\label{potential}V_{S}(r)=f(r)\frac{U(r)}{16r^2H(r)^2},
\end{equation}
where
\begin{equation}\label{pole-function}
H(r)=\lambda+X \frac{d(d+1)}{2}-Zd^2,
\end{equation}
\begin{widetext}
\begin{eqnarray}\nonumber
U(r)&=&\left(-d^3(d + 1)^2(d + 2)X^2+4d^2(d + 1)\left(d(d^2 + 6d - 4)Z + 3(d - 2)\lambda\right)X - 12d^5(3d - 2)Z^2 -\right.
\\\nonumber && \left. - 8d^2(11d^2 - 26d + 12)\lambda Z - 4(d - 2)(d - 4)\lambda^2\right)Y + d^4(d + 1)^2X^3 + d(d + 1) \left(-3d^2(5d^2 - 5d + 2) Z + \right.
\\\nonumber && \left. + 4(2d^2 - 3d + 4)\lambda + d(d - 2)(d - 4)(d + 1)\right)X^2 + 4d\left(d^2(4d^3 + 5d^2 - 10d + 4)Z^2 - \right.
\\\nonumber && \left. - \left(d(34 - 43d + 21d^2)\lambda + d^2(d + 1)(d^2 - 10d + 12)\right)Z - 3(d - 4)\lambda^2 - 3d(d + 1)(d - 2) \lambda\right)X -
\\\nonumber && - 4d^5(3d - 2)Z^3 + 4)\lambda+ 12d^2\left(2(-6d + 3d^2 + 4)\lambda + d^2(3d - 4)(d - 2)\right)Z^2 + 4(13d - 4)(d - 2)\lambda^2 Z +
\\\nonumber && + 8d^2(11d^2 - 18d + 4)\lambda Z + 16\lambda^3 + 4d(d + 2) \lambda^2,
\end{eqnarray}
\end{widetext}
Here we used $d=D-2$, $\lambda=(l+d)(l-1)$ and $l$ is the
multipole number.

We shall measure all the quantities in the units of the horizon
radius. Thus we define
\begin{equation}\label{horizon-units}
2M=1+Q^2-2\Lambda/d(d+1).
\end{equation}

In order to parameterize the cosmological constant, we introduce
the parameter $\rho=1/r_c<1$, where $r_c$ is the cosmological
horizon. The value $\rho=0$ corresponds to the pure
$D$-dimensional Reissner-Nordstr\"om black hole. Then, $\Lambda$
in terms of $Q$ and $\rho$ can be find from the equation
\begin{equation}\label{Lambda-parameterize}
f(r_c)=f(1/\rho)=0.
\end{equation}
Solving the equation (\ref{Lambda-parameterize}), one can find
\begin{equation}\label{Lambda-solution}
\Lambda=\frac{d(d+1)}{2}\frac{1-\rho^{1-d}+Q^2(1-\rho^{d-1})}{1-\rho^{-d-1}}.
\end{equation}

It is clear that for each $\Lambda$ there is some extremal charge
$Q_{ext}$. The value of $Q_{ext}$ can be found from
\begin{equation}\label{extremal-charge}
f'(1)=0.
\end{equation}
For any charge $Q>Q_{ext}$, the definition (\ref{horizon-units})
is not correct because in this case, the real event horizon has
radius greater than one.

Substituting (\ref{Lambda-solution}) into (\ref{extremal-charge}),
one can find
\begin{equation}\label{extremal-charge-found}
Q_{ext}^2=\frac{(d-1)(1-\rho^{d+1})-\rho^2(d+1)(1-\rho^{d-1})}{(d-1)(1-\rho^{d+1})-\rho^{d+1}(d+1)(1-\rho^{d-1})}.
\end{equation}

Within this paper we shall use $\rho$ and $q=Q/Q_{ext}$ as the
black hole parameters. Note that the formulas
(\ref{extremal-charge-found}) and (\ref{Lambda-solution}) allow us
to find $Q$ and $\Lambda$ in terms of $0\leq\rho<1$ and $0\leq
q<1$. In the particular asymptotically flat case ($\Lambda=0$), we
have $Q_{ext}=1$ and, therefore, $Q=q$.

\section{Numerical methods for frequency and time domains}\label{numericalmethods::sec}

\subsection{Frequency domain}
\begin{figure}
\resizebox{\linewidth}{!}{\includegraphics*{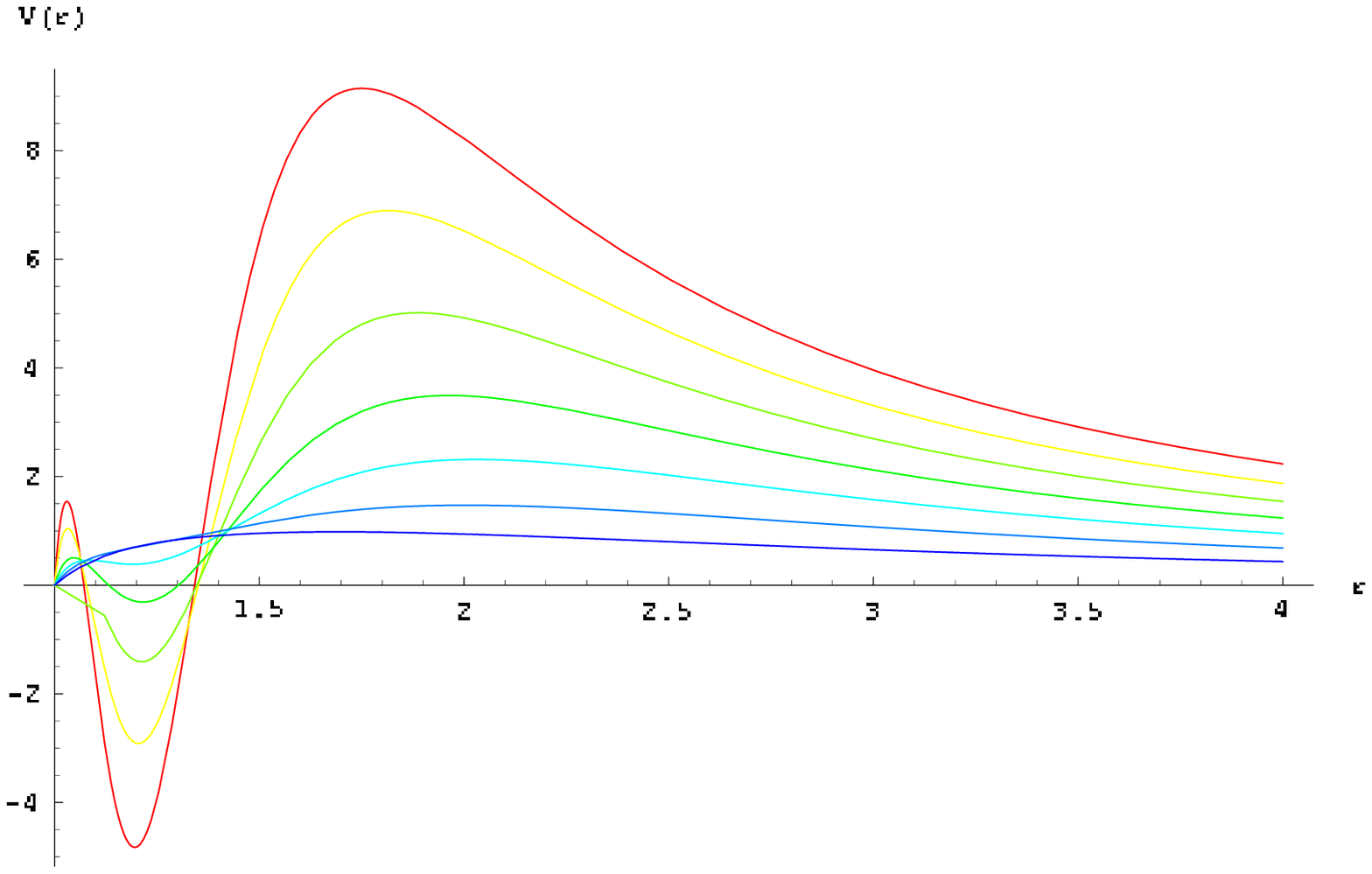}}
\caption{Effective potentials for gravitational perturbations of scalar type, $D=5$ (blue)\ldots $D=11$ (red) ($l=2$, $Q=0$, $\Lambda=0$). For higher $D$ both the peak and the negative gap of the potential increase.}
\end{figure}
\begin{figure}
\resizebox{\linewidth}{!}{\includegraphics*{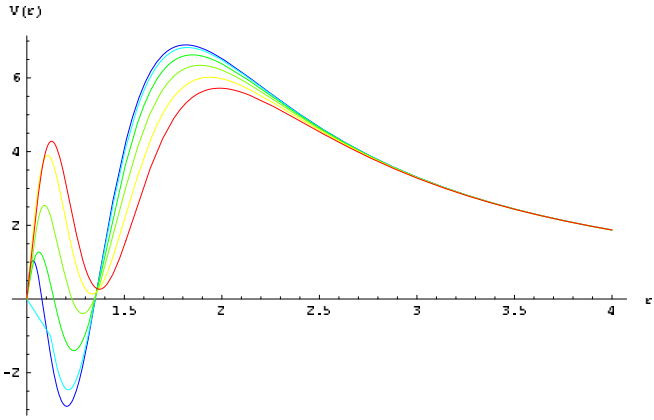}}
\caption{Effective potentials for gravitational perturbations of scalar type, $Q=0$ (blue), $Q=0.2$ (light blue), $Q=0.4$ (green), $Q=0.6$ (light green), $Q=0.8$ (yellow), $Q=0.98$ (red) ($l=2$, $D=10$, $\Lambda=0$). Increasing of the charge $Q$ cause the negative gap to move upwards. For some $Q$ the minimum of the potential becomes positive.}
\end{figure}
\begin{figure}
\resizebox{\linewidth}{!}{\includegraphics*{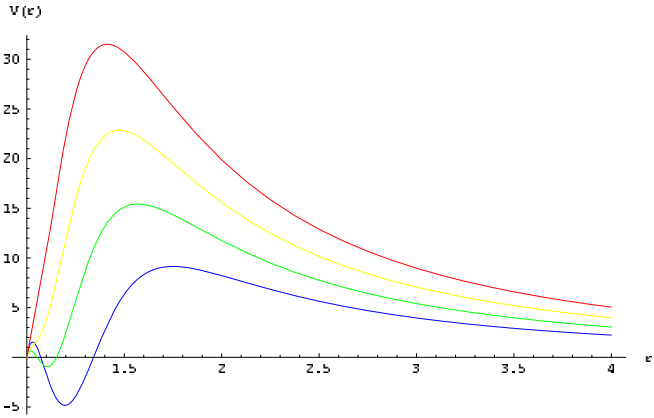}}
\caption{Effective potentials for gravitational perturbations of scalar type, $l=2,3,4,5$ (blue, green, yellow, red) ($D=11$, $Q=0$, $\Lambda=0$). For high multipole numbers the potential minimum disappears. Thus for $8\leq D\leq10$ the negative gap exists only for the lowest multipole number $l=2$.}
\end{figure}

\begin{figure}
\resizebox{\linewidth}{!}{\includegraphics*{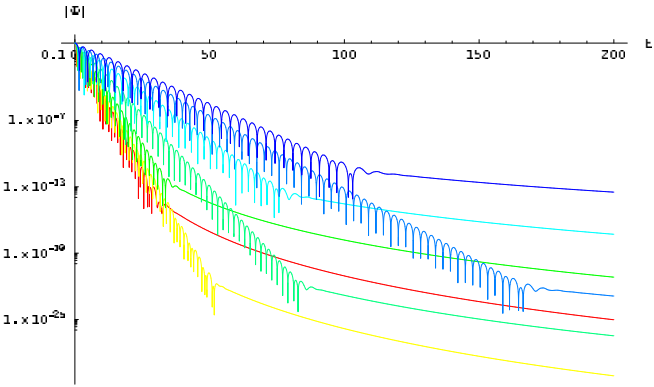}}
\caption{Time-domain profiles for gravitational perturbations of scalar type ($Q=0$, $\Lambda=0$) for $D=5$ (blue)\ldots $D=11$ (red) at the same point $r=2$. Profile for higher $D$ decays quicker.}
\end{figure}

\begin{figure}
\resizebox{\linewidth}{!}{\includegraphics*{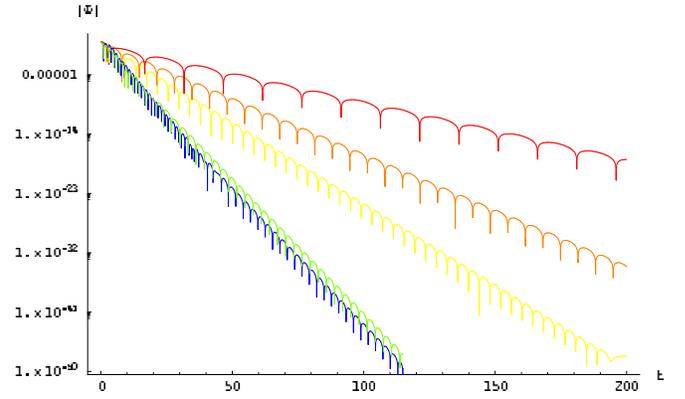}}
\caption{Time-domain profiles for gravitational perturbations of scalar type ($Q=0$, $D=11$) for $\rho=0.3$ (blue), $\rho=0.5$ (green), $\rho=0.7$ (yellow), $\rho=0.8$ (orange), $\rho=0.9$ (red). Profile for higher $\rho$ decays slower.}
\end{figure}

\begin{figure}
\resizebox{\linewidth}{!}{\includegraphics*{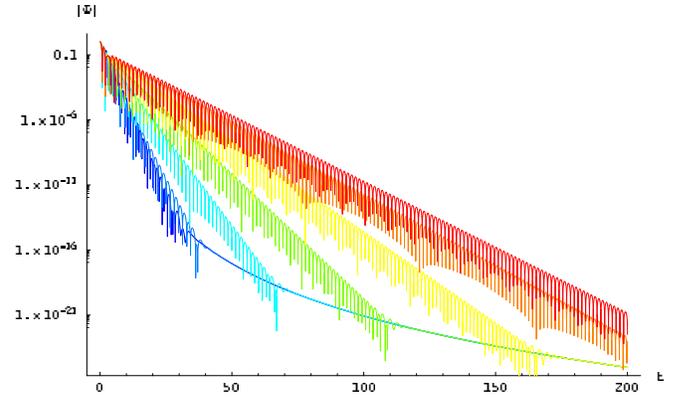}}
\caption{Time-domain profiles for gravitational perturbations of scalar type ($\Lambda=0$, $D=11$) for $Q=0$ (dark blue), $Q=0.5$ (blue), $Q=0.6$ (light blue), $Q=0.7$ (green), $Q=0.8$ (yellow), $Q=0.9$ (orange), $Q=0.98$ (red). Profile for higher $Q$ decays slower.}
\end{figure}

\begin{figure}
\resizebox{\linewidth}{!}{\includegraphics*{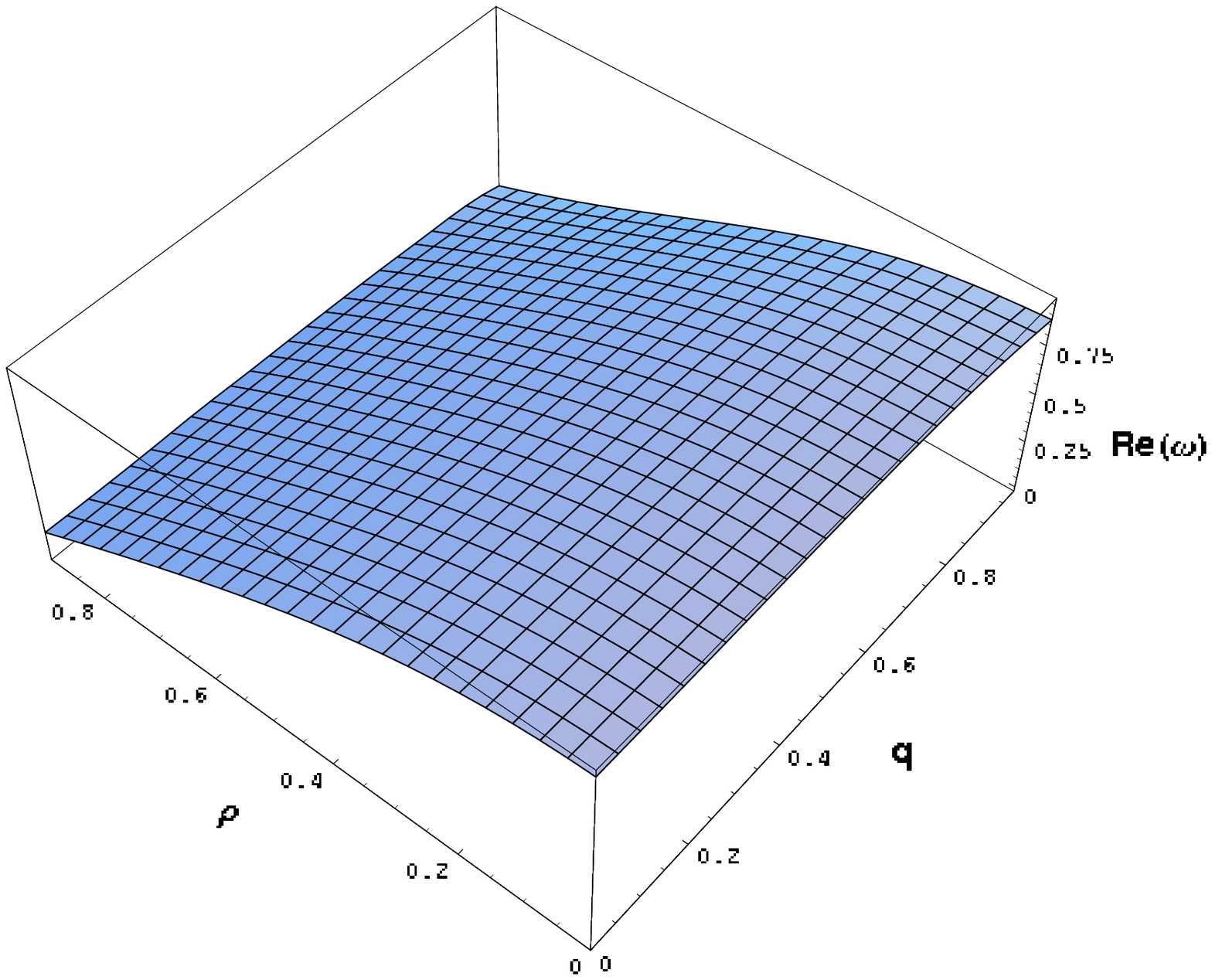}}
\caption{Real and imaginary part of the fundamental quasinormal frequency of gravitational perturbations of scalar type as a function of $q$ and $\rho$ ($D=5$, $l=2$).}\label{QNM.D=5.fig}
\resizebox{\linewidth}{!}{\includegraphics*{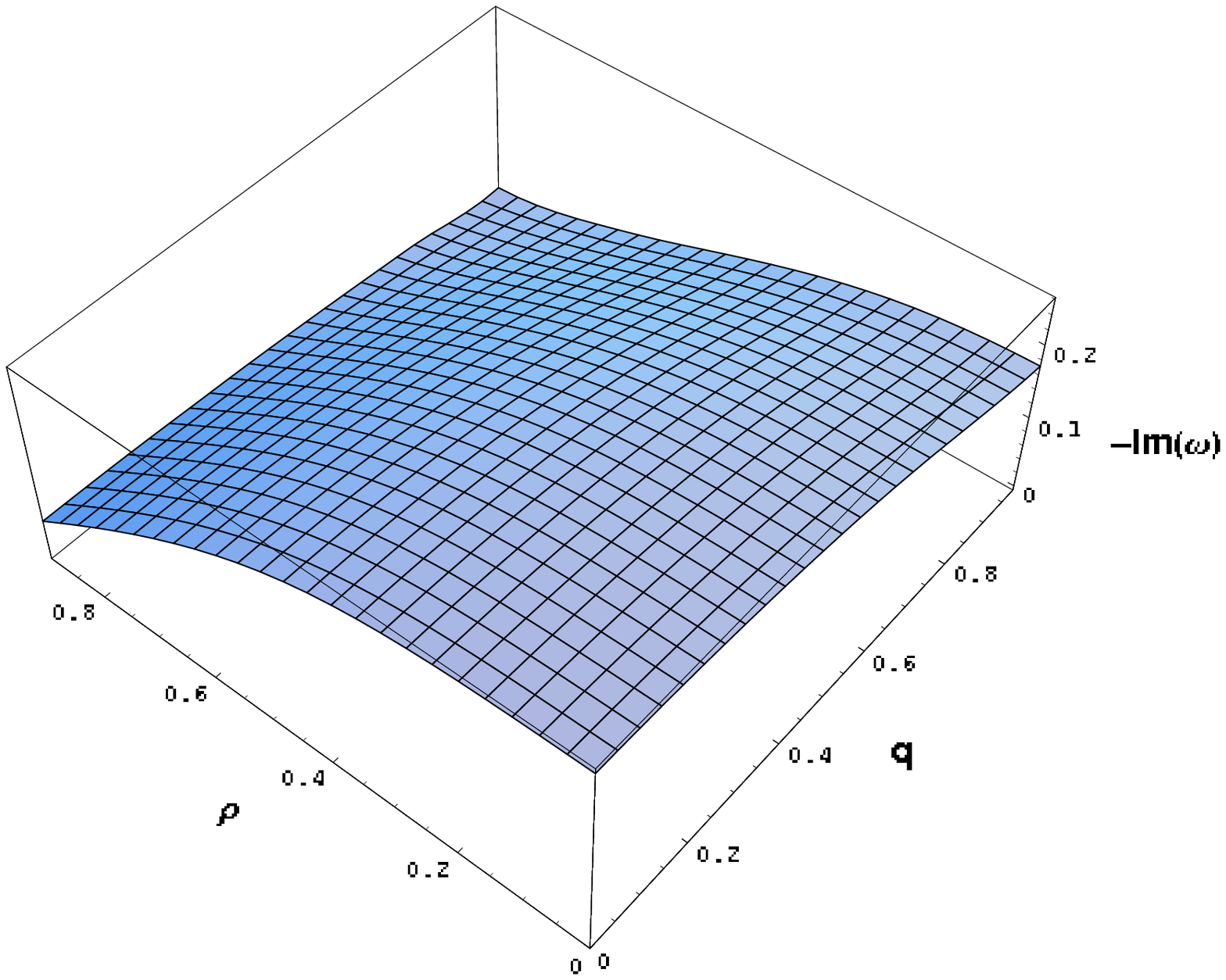}}
\end{figure}

\begin{figure}
\resizebox{\linewidth}{!}{\includegraphics*{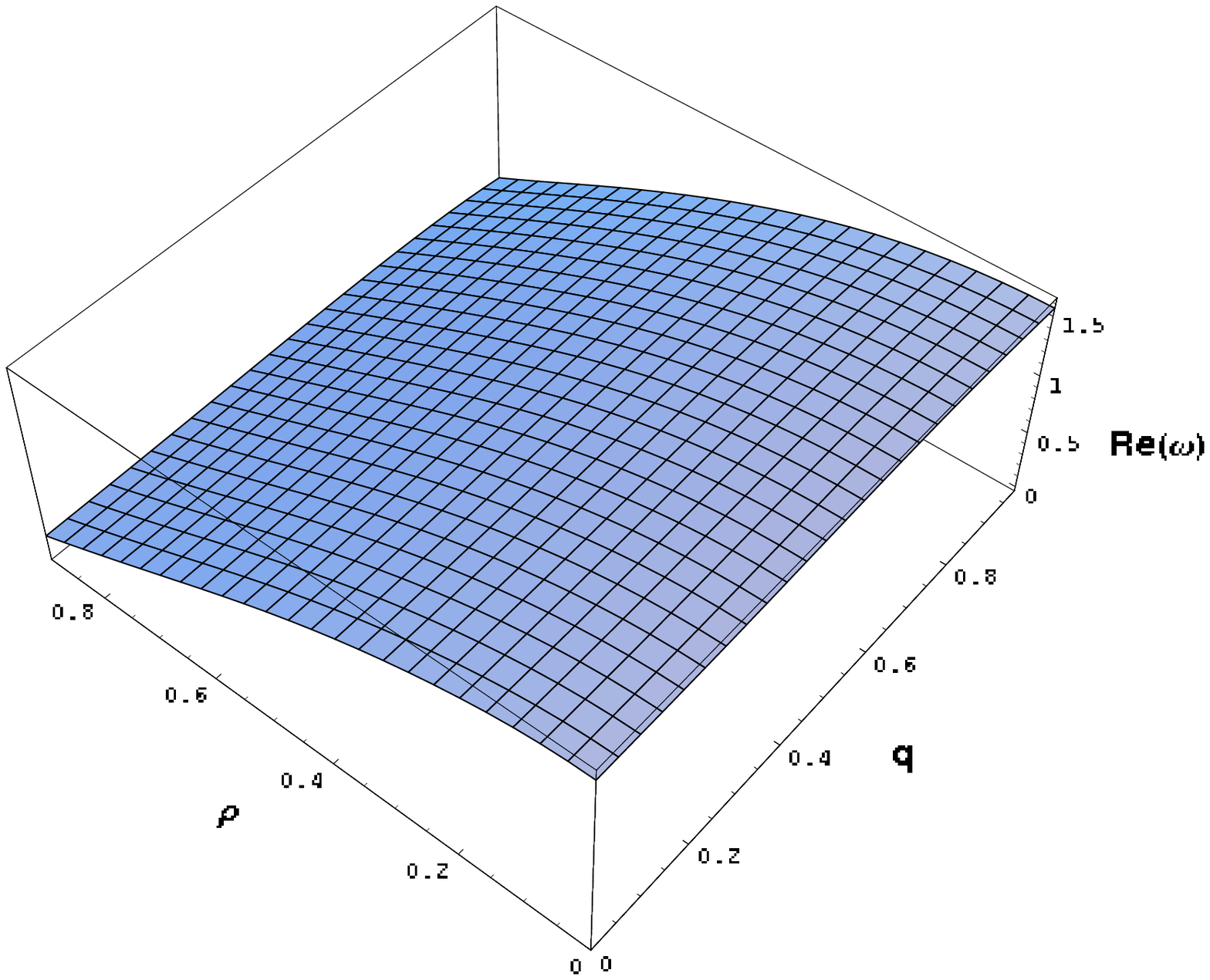}}
\caption{Real and imaginary part of the fundamental quasinormal frequency of gravitational perturbations of scalar type as a function of $q$ and $\rho$ ($D=8$, $l=2$).}\label{QNM.D=8.fig}
\resizebox{\linewidth}{!}{\includegraphics*{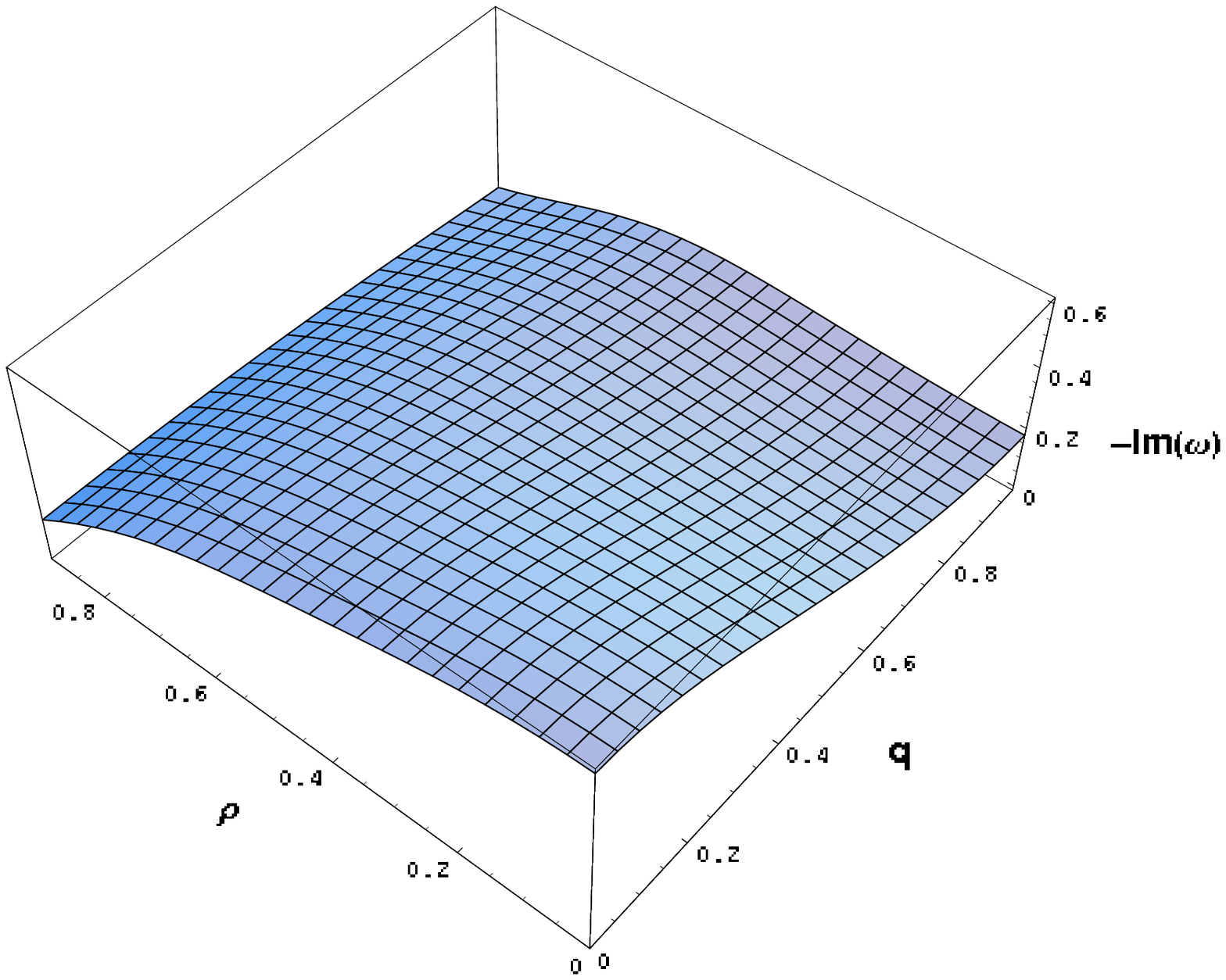}}
\end{figure}

\begin{figure}
\resizebox{\linewidth}{!}{\includegraphics*{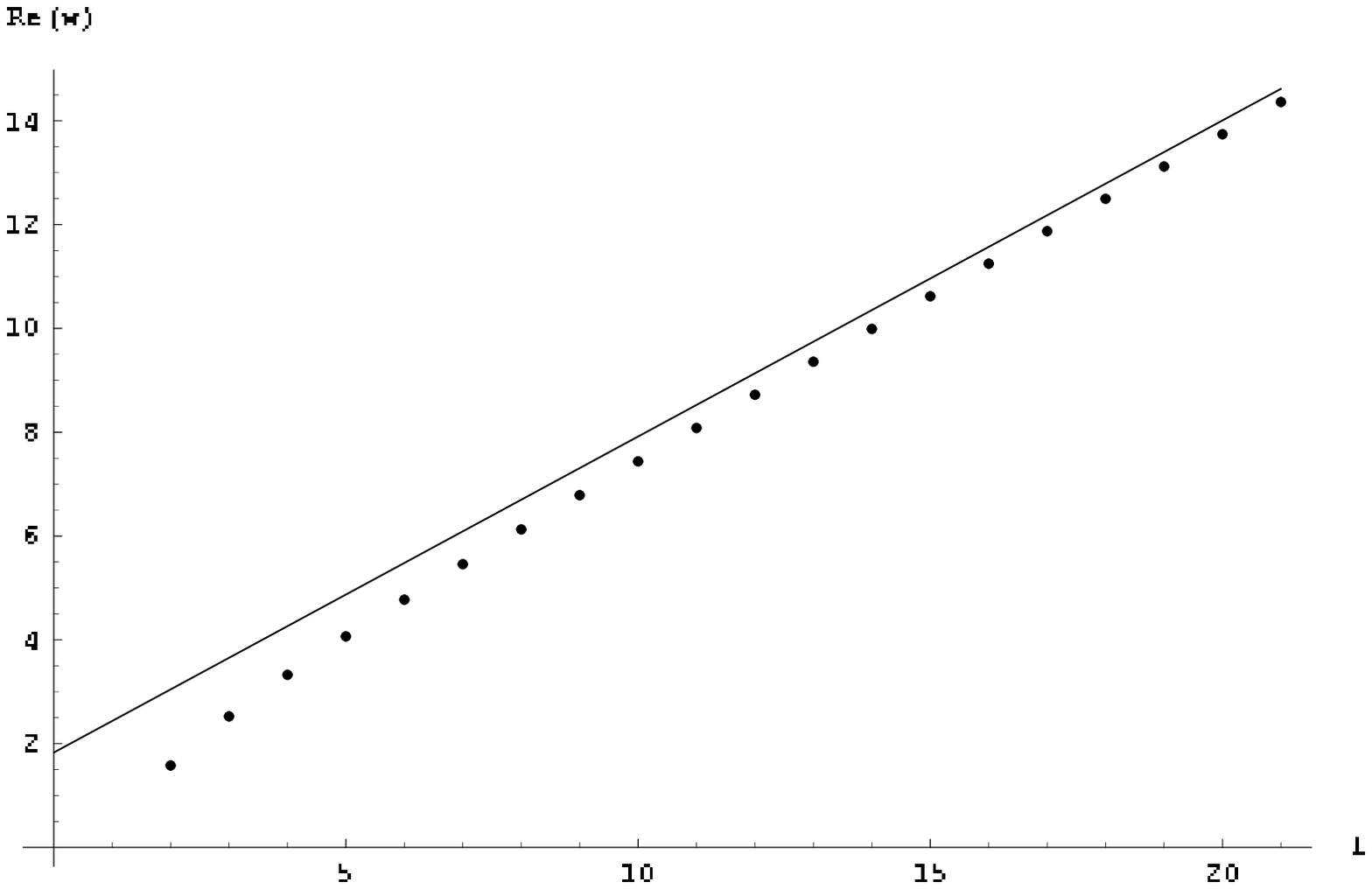}}
\caption{Real and imaginary part of the fundamental quasinormal frequency of gravitational perturbations of scalar type as a function of $l$ ($D=9$, $q=0.5$, $\rho=0.3$). The solid line represents high multipole WKB formula.}\label{QNM.D=9.fig}
\resizebox{\linewidth}{!}{\includegraphics*{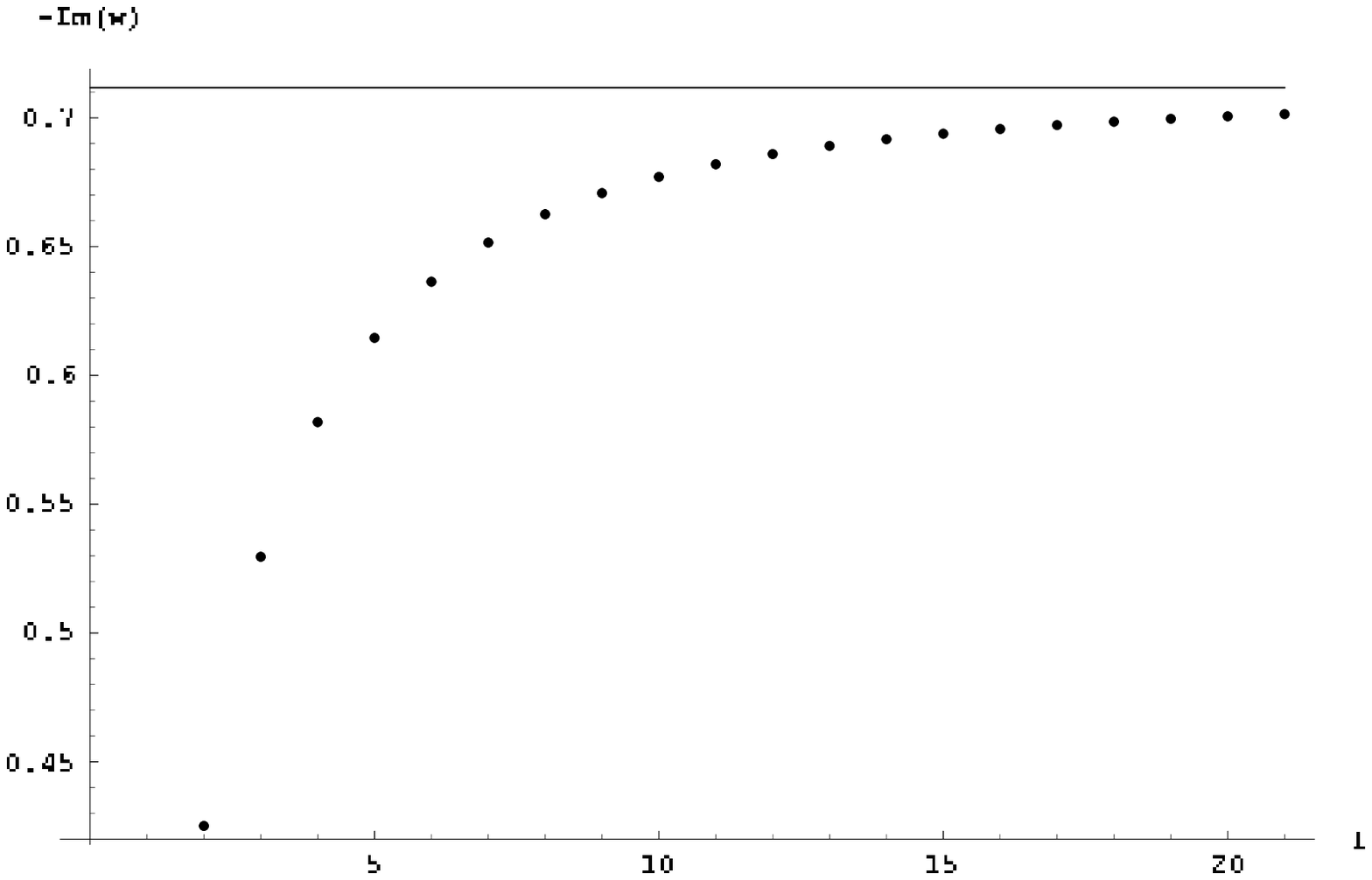}}
\end{figure}

By definition, the quasi-normal modes are eigenvalues of $\omega$
with the boundary conditions which correspond to the outgoing wave
at the cosmological horizon (or spatial infinity when $\rho=0$)
and the ingoing wave at the black hole horizon.

In order to find quasi-normal spectrum for the considered types of
perturbations, we shall use the continued fraction method
\cite{Leaver:1985ax}. The equation (\ref{wavelike}) has a different
kind of singularity at $r=1/\rho$: if $\rho=0$, the singularity is
irregular at infinity as usually for the Schwarzschild case
$$\Psi(r\rightarrow\infty)\propto e^{\imo\omega r},$$
while for $\rho>0$ the point of the cosmological horizon is a
regular singularity \cite{Yoshida:2003zz}
$$\Psi\left(r\rightarrow\frac{1}{\rho}\right)\propto\left(\rho -\frac{1}{r}\right)^{\imo\omega/f'\left(\frac{1}{\rho}\right)}.$$
It turns out, that we can simply exclude this singularity by the
introducing of the new function
$$\Phi=e^{-\imo\omega  r^*}\Psi,$$
where $\Phi$ satisfies
\begin{equation}\label{phi-equation}
\frac{d^2\Phi}{dr^{*2}}+2\imo\omega\frac{d\Phi}{dr^*}-V\Phi=0.
\end{equation}
In order to satisfy the QNM boundary conditions,  the function
$\Phi$ has to be convergent at $r=1/\rho$. Thus, the only valuable
singularity remains at $r=1$. Taking into account that there is
only the ingoing wave at the event horizon, one can find
$$\Phi(r\rightarrow1)\propto(r-1)^{-2\imo\omega/f'(1)}.$$
Finally, we obtain
\begin{equation}\label{nosingularity}
\Phi(r)=\left(1-\frac{1}{r}\right)^{\frac{-2\imo\omega}{f'(1)}}y(r),
\end{equation}
where y(r) is convergent on $r\in[1,1/\rho]$. The appropriate
Frobenius series for $y(r)$ are
\begin{equation}\label{Frobenius}
y(r)=\sum_{i=0}^\infty
a_n\left(\frac{r-1}{r-R}\frac{1-R\rho}{1-\rho}\right)^n,
\end{equation}
where $R<1$ is an {\it arbitrary} parameter.

The expansion (\ref{Frobenius}) is valid if all singular
points (except those on the event and the cosmological horizons)
satisfy inequality
\begin{equation}\label{singular-inequality}
\left|\frac{r-1}{r-R}\frac{1-R\rho}{1-\rho}\right|>1.
\end{equation}
Since for the effective potential of the gravitational
perturbations of vector and tensor types the singular points are
1) $r=0$ and 2) the solutions of the equation $f(r)=0$, we can
always find such $0\leq R<1$ to satisfy inequality
(\ref{singular-inequality}) for all the additional singular
points. The latter follows from the definition of the horizons,
which are the two larger positive solutions of the equation
$f(r)=0$.

After $R$ is fixed, following \cite{Zhidenko:2006rs}, one can find
the recurrence relation for the coefficients $a_n$ in
(\ref{Frobenius}) and use the continued fraction method with the
generalized Nollert improvement \cite{Nollert}. It turns out, that the
convergence of the method becomes worse when $R\rightarrow 1$.
Thus one has to choose the minimal value of $R$ that satisfies the
inequality (\ref{singular-inequality}) for all singular points,
except those at the event and the cosmological horizons. For
$D\leq9$ this point is $R=r_-$, which is the remaining positive
root of $f(r)$ (or $R=0$ if $f(r)$ has no more positive root), and
the continued fraction method converges without the generalized
Nollert improvement. For larger $D$ such $R$ is, generally, not a
singular point of the equation (\ref{phi-equation}) and the
convergence of the method is provided by the procedure
\cite{Zhidenko:2006rs}. Alternatively, for $D>9$ one can continue
the Frobenius series through some midpoints. This method was used
in \cite{Rostworowski:2006bp} and our results are completely in
agreement with it.

For the scalar-type gravitational perturbations, the equation
(\ref{phi-equation}) has additional singularities, which are the
roots of $H(r)$. Thus, in higher dimensions $D\geq7$, for some
values of $\rho$, $q$ and $l$ there is no such $R<1$, which
satisfies the inequality (\ref{singular-inequality}) for all the
singular points. Therefore, for those cases one should use only
the technique developed in \cite{Rostworowski:2006bp}.
Practically, the continued fraction coefficients appear to be so
complicated, that we are unable to compute QNMs with this method
during reasonable time, even for low dimensions, except the case
when $\rho=0$.

\subsection{Time domain}
Alternatively, we shall use the time-domain analysis based on the
standard integration scheme for the wave-like equations (\ref{hyperbolic}),
which is described for instance in \cite{Price-Pullin}. This approach has
no limitations of the parameters for the scalar-type
perturbations. Moreover, it appears to be quicker for the
perturbations of scalar type, when $D$ is large or $\rho\neq0$.
Since we can see the real signal which contains all the
frequencies, this method also provides the most direct evidence of
the black hole stability. It allows also to find out asymptotical
tails.

In detail, we applied a numerical characteristic integration
scheme,that uses the light-cone variables $u = t - r_\star$ and $v
= t + r_\star$. In the characteristic initial value problem,
initial data are specified on the two null surfaces $u = u_{0}$
and $v = v_{0}$. The discretization scheme we used, is
\begin{eqnarray}
\lefteqn{\Psi(N) = \Psi(W) + \Psi(E) -
\Psi(S) }  \nonumber \\
& & \mbox{} - \Delta^2\frac{V(W)\Psi(W) + V(E)\Psi(E)}{8} +
\mathcal{O}(\Delta^4)   \ ,
\label{d-uv-eq}
\end{eqnarray}
where we have used the following definitions for the points: $N =
(u +
\Delta, v + \Delta)$, $W = (u + \Delta, v)$, $E = (u, v + \Delta)$
and $S = (u,v)$.

\section{Evolution of gravitational perturbations}

\subsection{Scalar type of gravitational perturbations}
\begin{table*}
\caption{
Fundamental modes ($l=2$, $\Lambda=0$) of gravitational perturbations of scalar type calculated by the continued fraction method (the first value) and by the time-domain simulation (the second value). For higher $D$ the behavior of the time-domain profile is very irregular. Thus QNMs finding with high accuracy is more complicated.}\label{scalarQQNMs}
\begin{tabular}{|l|c|c|c|c|c|c|}
\hline
$Q$&$D=5$&$D=6$&$D=7$&$D=8$&$D=9$&$D=10$\\
\hline
$0   $&$0.94774-0.25609\imo$&$1.13690-0.30358\imo$&$1.33916-0.40086\imo$&$\!\!-$&$\!\!-$&$\!\!-$\\
      &$0.94773-0.25609\imo$&$1.13690-0.30357\imo$&$1.33917-0.40084\imo$&$1.56390-0.60312\imo$&$1.99898-0.86335\imo$&$2.45~-~0.98\imo$\\
$0.1 $&$0.94842-0.25523\imo$&$1.13830-0.30117\imo$&$1.34131-0.39489\imo$&$\!\!-$&$\!\!-$&$\!\!-$\\
      &$0.94841-0.25522\imo$&$1.13830-0.30117\imo$&$1.34132-0.39488\imo$&$1.56327-0.58781\imo$&$1.98516-0.85771\imo$&$2.45~-~0.97\imo$\\
$0.2 $&$0.95032-0.25263\imo$&$1.14225-0.29422\imo$&$1.34745-0.37794\imo$&$\!\!-$&$\!\!-$&$\!\!-$\\
      &$0.95032-0.25262\imo$&$1.14225-0.29422\imo$&$1.34745-0.37793\imo$&$1.56447-0.54593\imo$&$1.93834-0.83840\imo$&$2.41~-~0.97\imo$\\
$0.3 $&$0.95304-0.24830\imo$&$1.14802-0.28342\imo$&$1.35661-0.35246\imo$&$\!\!-$&$\!\!-$&$\!\!-$\\
      &$0.95303-0.24829\imo$&$1.14802-0.28342\imo$&$1.35661-0.35246\imo$&$1.57120-0.48765\imo$&$1.83938-0.76949\imo$&$2.34~-~0.96\imo$\\
$0.4 $&$0.95585-0.24218\imo$&$1.15446-0.26977\imo$&$1.36721-0.32163\imo$&$\!\!-$&$\!\!-$&$\!\!-$\\
      &$0.95583-0.24217\imo$&$1.15447-0.26976\imo$&$1.36721-0.32163\imo$&$1.58264-0.42318\imo$&$1.79707-0.60660\imo$&$2.14~-~0.94\imo$\\
$0.5 $&$0.95762-0.23407\imo$&$1.16006-0.25436\imo$&$1.37715-0.28865\imo$&$\!\!-$&$\!\!-$&$\!\!-$\\
      &$0.95760-0.23405\imo$&$1.16006-0.25435\imo$&$1.37715-0.28865\imo$&$1.59530-0.35950\imo$&$1.80551-0.47879\imo$&$1.985-0.674\imo$\\
$0.6 $&$0.95666-0.22363\imo$&$1.16290-0.23816\imo$&$1.38398-0.25627\imo$&$\!\!-$&$\!\!-$&$\!\!-$\\
      &$0.95662-0.22361\imo$&$1.16289-0.23816\imo$&$1.38398-0.25627\imo$&$1.60536-0.30104\imo$&$1.81782-0.37688\imo$&$2.011-0.486\imo$\\
$0.7 $&$0.95033-0.21055\imo$&$1.16059-0.22197\imo$&$1.38498-0.22676\imo$&$\!\!-$&$\!\!-$&$\!\!-$\\
      &$0.95027-0.21052\imo$&$1.16060-0.22197\imo$&$1.38498-0.22675\imo$&$1.60921-0.25088\imo$&$1.82430-0.29638\imo$&$2.024-0.362\imo$\\
$0.8 $&$0.93502-0.19538\imo$&$1.15025-0.20660\imo$&$\!\!-$&$\!\!-$&$\!\!-$&$\!\!-$\\
      &$0.93490-0.19535\imo$&$1.15026-0.20659\imo$&$1.37723-0.20212\imo$&$1.60359-0.21163\imo$&$1.82052-0.23687\imo$&$2.023-0.275\imo$\\
$0.9 $&$0.90758-0.18077\imo$&$1.12912-0.19364\imo$&$\!\!-$&$\!\!-$&$\!\!-$&$\!\!-$\\
      &$0.90745-0.18075\imo$&$1.12913-0.19363\imo$&$1.35858-0.18465\imo$&$1.58656-0.18609\imo$&$1.80461-0.20010\imo$&$2.009-0.224\imo$\\
$0.98$&$0.87720-0.17177\imo$&$1.10420-0.18636\imo$&$\!\!-$&$\!\!-$&$\!\!-$&$\!\!-$\\
      &$0.87707-0.17175\imo$&$1.10429-0.18641\imo$&$1.33609-0.17704\imo$&$1.56549-0.17657\imo$&$1.78451-0.18755\imo$&$1.990-0.208\imo$\\
\hline
\end{tabular}
\end{table*}

\begin{table*}
\caption{Fundamental modes ($l=2$, $Q=0$) of gravitational perturbations of scalar type calculated by the time-domain simulation. For higher $D$ the behavior of the time-domain profile is very irregular. Thus QNMs finding with high accuracy is more complicated.}\label{scalarRQNMs}
\begin{tabular}{|l|c|c|c|c|c|c|}
\hline
$\rho$&$D=5$&$D=6$&$D=7$&$D=8$&$D=9$&$D=10$\\
\hline
$  0$&$0.94774-0.25609\imo$&$1.13690-0.30358\imo$&$1.33916-0.40086\imo$&$1.56390-0.60312\imo$&$1.99898-0.86335\imo$&$2.45~-~0.98\imo$\\
$0.1$&$0.93336-0.25576\imo$&$1.12050-0.30461\imo$&$1.31868-0.39970\imo$&$1.53691-0.59412\imo$&$1.95602-0.86639\imo$&$2.41~-~0.99\imo$\\
$0.2$&$0.89204-0.25366\imo$&$1.07169-0.30664\imo$&$1.25828-0.39567\imo$&$1.45301-0.56754\imo$&$1.81939-0.87158\imo$&$2.28~-~1.00\imo$\\
$0.3$&$0.82569-0.24749\imo$&$0.99203-0.30681\imo$&$1.16108-0.38781\imo$&$1.33119-0.52832\imo$&$1.56112-0.84743\imo$&$2.04~-~1.01\imo$\\
$0.4$&$0.73774-0.23477\imo$&$0.88551-0.30073\imo$&$1.03263-0.37458\imo$&$1.17909-0.48504\imo$&$1.31893-0.67923\imo$&$1.70~-~1.01\imo$\\
$0.5$&$0.63289-0.21293\imo$&$0.75787-0.28372\imo$&$0.87872-0.35286\imo$&$0.99978-0.44032\imo$&$1.11688-0.56723\imo$&$1.209-0.801\imo$\\
$0.6$&$0.51662-0.18156\imo$&$0.61591-0.25212\imo$&$0.70774-0.31741\imo$&$0.79938-0.38928\imo$&$0.89080-0.47868\imo$&$0.978-0.601\imo$\\
$0.7$&$0.39236-0.14257\imo$&$0.46597-0.20493\imo$&$0.52888-0.26326\imo$&$0.58941-0.32283\imo$&$0.64959-0.38903\imo$&$0.710-0.468\imo$\\
$0.8$&$0.26341-0.09797\imo$&$0.31181-0.14428\imo$&$0.34978-0.18904\imo$&$0.38262-0.23377\imo$&$0.41294-0.28050\imo$&$0.442-0.332\imo$\\
$0.9$&$0.12993-0.04982\imo$&$0.15447-0.07427\imo$&$0.17045-0.09844\imo$&$0.18471-0.12283\imo$&$0.19478-0.14745\imo$&$0.203-0.173\imo$\\
\hline
\end{tabular}
\end{table*}

It is essential for our computations that the effective potential for charged black holes have the negative gap which cannot be "removed" by S-deformation \cite{Ishibashi:2003ap,Kodama:2003kk,Kodama:2003jz} for $D\geq 8$. This gives us the cases for suspecting instability.  The fundamental quasinormal modes for scalar type of gravitational perturbations computed with Frobenius and time-domain techniques are given in Tables \ref{scalarQQNMs}, \ref{scalarRQNMs}. For $D=5, 6, 7$ Reissner-Nordstr\"om black holes ($\Lambda =0$), the real parts of QNMs increase with charge until some maximum, then decrease.  For $D \geq 8$ $Re \omega$, for small $Q$, decreases as a function of $Q$, then increases and reaches some maximum and then, again decreases for close to the extremal $Q$. The imaginary part of the frequencies as a function of charge $Q$ monotonically decreases for all $D$ (Table \ref{scalarQQNMs}). Note that in table \ref{scalarQQNMs} the calculations of QNMs were done for $D=5, 6, 7$ with both time-domain and Frobenius approaches. Yet, for higher $D$, the Frobenius method needs considerable computer time for converging. The time-domain picture of the decay for black holes with high $D=10, 11, \ldots$ are different from lower dimensional cases: if for $D=5, 6, \ldots9$ we see always regular picture of time-domain decay (Fig. 4), for $D= 10, 11$, and for small $Q$ and $\Lambda$ the ringing is more irregular (in fact, the amplitude is irregular, that probably means the superposition of a few dominating modes) and quickly changes into tail behavior (Fig.5, Fig. 6). The time domain picture becomes again regular at larger $Q$ and $\Lambda$ (Fig.5, Fig. 6).

The fundamental modes for Schwarzschild -de Sitter black holes ($Q=0$) are given in Table \ref{scalarRQNMs}. The real oscillation
frequency usually monotonically decreases as a function of $\Lambda$-term. Note that literature on QNMs of the scalar type of gravitational perturbations in $D>4$ is very poor \cite{Konoplya:2003dd,cardosoJHEP}. Our computations show perfect coincidence with Frobenius method data of \cite{cardosoJHEP} and very good coincidence with the WKB results of \cite{Konoplya:2003dd} for scalar type of gravitational perturbations. Alternative approach for finding gravitational quasinormal modes is to consider the head-on collisions for higher
dimensional theories  \cite{Yoshino:2005ps,Yoshino:2006kc}. The comparison with our accurate numerical values of QNMs gives very good agreement for not large $D$: Thus for $D=5$ and $6$, we have $\omega = 0.94773 - 0.25609\imo$, $\omega = 1.13690 - 0.30357\imo$ respectively, while paper \cite{Yoshino:2005ps} give the values $\omega = 0.947 - 0.256\imo$, $\omega = 1.139 - 0.305\imo$, i.e. the error in \cite{Yoshino:2005ps} is indeed very small and is less than $1$ percent. For $D=9, 10, 11$ the method in \cite{Yoshino:2005ps} shows superposition of the two dominating modes, one of which is very close to that obtained here. For instance for $D=4$, they have $\omega = 2.47 -0.99\imo$, while we obtained with time-domain approach $\omega = 2.45 - 0.98\imo$. This small difference between our results is quite natural because different approximations were used: we considered the regime of small perturbations and in \cite{Yoshino:2005ps} a close limit approximation was used. Therefore we conclude that the close limit approximation with Brill-Lindquist initial data is indeed a very accurate approach.

When both parameters $Q$ and $\Lambda$ are not vanishing, the quasinormal frequencies are again damping. An example of behavior of real and imaginary of $\omega$ as a function of $q$ and $\rho$ (which parameterize all the values of $Q$ and $\Lambda$) is given in Fig \ref{QNM.D=5.fig} for $D=5$ case and Fig \ref{QNM.D=8.fig} for $D=8$ case.

At large multi-pole numbers $\ell$, the WKB approach \cite{WKB}
becomes very accurate and one can obtain usually an analytical
formula for the QN frequencies. For our case of
Reissner-Nordstr\"om-de Sitter black holes, a cumbersome form of the
effective potential with a lot of parameters makes it impossible
to find a short QNM formula, which could be easily interpreted for
general values of $M$, $Q$, $\Lambda$, $D$. Yet, if we fix all the
above parameters the WKB formula in the lowest order gives, for
example for $D=9$, $Q=0.5 Q_{ext}$, $\rho =0.3$ (see Fig \ref{QNM.D=9.fig}):

\begin{equation}
\omega = 0.60897(\ell+3)-1.42338(n+1/2)\imo, \quad \ell \rightarrow \infty
\end{equation}

Here we can see that in a similar fashion with a four-dimensional
case, the real oscillation frequency is proportional to $\ell$ and
the damping rate is approaching some constant value. Therefore one
should not expect any instability from the regime of high $\ell$.
Indeed, higher $\ell$ just raise the potential barrier and serve
as a stabilizing factor: perturbations with higher $\ell$ are more
stable than perturbations with lower $\ell$.

Finally, summarizing the extensive search for quasinormal modes
for scalar type of gravitational perturbations, we conclude that
the response to the external perturbations of the
Reissner-Nordstr\"om-de Sitter black holes are represented by
damping oscillations and therefore {\it for all values of charge
and $\Lambda$-term in $D=5, 6, \ldots11$ space-time dimensions, the
Reissner-Nordstr\"om-de Sitter black holes are stable}.

\subsection{Vector and tensor types of gravitational perturbations}

\begin{table*}
\caption{Fundamental modes ($l=2$, $\Lambda=0$) of gravitational perturbations of vector($"+"$) type calculated by the time-domain simulation.}\label{vector+QQNMs}
\begin{tabular}{|l|c|c|c|c|c|c|}
\hline
$Q$&$D=5$&$D=6$&$D=7$&$D=8$&$D=9$&$D=10$\\
\hline
$0   $&$1.46854-0.35242\imo$&$1.97832-0.49641\imo$&$2.47008-0.62640\imo$&$2.95282-0.74529\imo$&$3.43062-0.85514\imo$&$3.90559-0.95743\imo$\\
$0.1 $&$1.47080-0.35064\imo$&$1.98180-0.49437\imo$&$2.47403-0.62391\imo$&$2.95761-0.74270\imo$&$3.43587-0.85244\imo$&$3.91135-0.95445\imo$\\
$0.2 $&$1.47742-0.34551\imo$&$1.99114-0.48814\imo$&$2.48556-0.61683\imo$&$2.97099-0.73491\imo$&$3.45092-0.84399\imo$&$3.92777-0.94560\imo$\\
$0.3 $&$1.48547-0.33667\imo$&$2.00371-0.47750\imo$&$2.50181-0.60482\imo$&$2.99037-0.72177\imo$&$3.47300-0.82999\imo$&$3.95228-0.93076\imo$\\
$0.4 $&$1.49155-0.32391\imo$&$2.01585-0.46226\imo$&$2.51897-0.58770\imo$&$3.01180-0.70309\imo$&$3.49819-0.81005\imo$&$3.98080-0.90983\imo$\\
$0.5 $&$1.49210-0.30723\imo$&$2.02337-0.44241\imo$&$2.53244-0.56546\imo$&$3.03033-0.67894\imo$&$3.52125-0.78433\imo$&$4.00792-0.88286\imo$\\
$0.6 $&$1.48353-0.28708\imo$&$2.02189-0.41860\imo$&$2.53713-0.53882\imo$&$3.04036-0.65008\imo$&$3.53610-0.75362\imo$&$4.02713-0.85065\imo$\\
$0.7 $&$1.46249-0.26477\imo$&$2.00727-0.39257\imo$&$2.52816-0.50988\imo$&$3.03633-0.61882\imo$&$3.53655-0.72040\imo$&$4.03167-0.81580\imo$\\
$0.8 $&$1.42669-0.24279\imo$&$1.97680-0.36751\imo$&$2.50236-0.48241\imo$&$3.01453-0.58938\imo$&$3.51835-0.68929\imo$&$4.01676-0.78326\imo$\\
$0.9 $&$1.37633-0.22431\imo$&$1.93086-0.34705\imo$&$2.46010-0.46053\imo$&$2.97522-0.56631\imo$&$3.48165-0.66523\imo$&$3.98235-0.75837\imo$\\
$0.98$&$1.32775-0.21336\imo$&$1.88525-0.33525\imo$&$2.41682-0.44832\imo$&$2.93363-0.55377\imo$&$3.44148-0.65248\imo$&$3.94338-0.74548\imo$\\
\hline
\end{tabular}
\end{table*}

\begin{table*}
\caption{Fundamental modes ($l=2$, $\Lambda=0$) of gravitational perturbations of vector($"-"$) type calculated by the time-domain
simulation.}\label{vector-QQNMs}
\begin{tabular}{|l|c|c|c|c|c|c|}
\hline
$Q$&$D=5$&$D=6$&$D=7$&$D=8$&$D=9$&$D=10$\\
\hline
$0   $&$1.13400-0.32752\imo$&$1.52467-0.47412\imo$&$1.93447-0.61230\imo$&$2.35885-0.73871\imo$&$2.79285-0.85412\imo$&$3.23345-0.96025\imo$\\
$0.1 $&$1.12873-0.32542\imo$&$1.51974-0.47195\imo$&$1.93008-0.60995\imo$&$2.35455-0.73593\imo$&$2.78882-0.85126\imo$&$3.22923-0.95699\imo$\\
$0.2 $&$1.11323-0.31965\imo$&$1.50551-0.46553\imo$&$1.91694-0.60268\imo$&$2.34200-0.72777\imo$&$2.77645-0.84240\imo$&$3.21691-0.94756\imo$\\
$0.3 $&$1.08935-0.31042\imo$&$1.48324-0.45520\imo$&$1.89595-0.59098\imo$&$2.32163-0.71475\imo$&$2.75614-0.82835\imo$&$3.19652-0.93271\imo$\\
$0.4 $&$1.05895-0.29829\imo$&$1.45430-0.44154\imo$&$1.86800-0.57559\imo$&$2.29396-0.69786\imo$&$2.72819-0.81033\imo$&$3.16820-0.91389\imo$\\
$0.5 $&$1.02349-0.28402\imo$&$1.41985-0.42551\imo$&$1.83391-0.55781\imo$&$2.25963-0.67875\imo$&$2.69314-0.79031\imo$&$3.13248-0.89327\imo$\\
$0.6 $&$0.98411-0.26862\imo$&$1.38094-0.40854\imo$&$1.79473-0.53951\imo$&$2.21974-0.65967\imo$&$2.65228-0.77079\imo$&$3.09083-0.87359\imo$\\
$0.7 $&$0.94199-0.25339\imo$&$1.33903-0.39228\imo$&$1.75230-0.52260\imo$&$2.17659-0.64270\imo$&$2.60824-0.75390\imo$&$3.04653-0.85697\imo$\\
$0.8 $&$0.89871-0.23937\imo$&$1.29619-0.37763\imo$&$1.70919-0.50774\imo$&$2.13312-0.62808\imo$&$2.56415-0.73951\imo$&$3.00213-0.84311\imo$\\
$0.9 $&$0.85578-0.22668\imo$&$1.25410-0.36428\imo$&$1.66671-0.49423\imo$&$2.09047-0.61471\imo$&$2.52102-0.72627\imo$&$2.95869-0.83045\imo$\\
$0.98$&$0.82248-0.21747\imo$&$1.22123-0.35412\imo$&$1.63310-0.48428\imo$&$2.05718-0.60463\imo$&$2.48645-0.71566\imo$&$2.92469-0.82086\imo$\\
\hline
\end{tabular}
\end{table*}

\begin{table*}
\caption{Fundamental modes ($l=2$, $Q=0$) of gravitational perturbations of vector($"+"$) type calculated by the
continued fraction method (the first value) and by the time-domain simulation (the second value).}\label{vector+RQNMs}
\begin{tabular}{|l|c|c|c|c|c|c|}
\hline
$\rho$&$D=5$&$D=6$&$D=7$&$D=8$&$D=9$&$D=10$\\
\hline
$0  $&$1.46854-0.35243\imo$&$1.97830-0.49642\imo$&$2.47006-0.62642\imo$&$2.95278-0.74532\imo$&$3.43056-0.85518\imo$&$3.90550-0.95749\imo$\\
     &$1.46854-0.35242\imo$&$1.97832-0.49641\imo$&$2.47008-0.62640\imo$&$2.95282-0.74529\imo$&$3.43062-0.85514\imo$&$3.90559-0.95743\imo$\\
$0.1$&$1.44629-0.34752\imo$&$1.95277-0.49123\imo$&$2.44074-0.62113\imo$&$2.91936-0.74005\imo$&$3.39283-0.84998\imo$&$3.86332-0.95241\imo$\\
     &$1.44629-0.34751\imo$&$1.95278-0.49122\imo$&$2.44076-0.62112\imo$&$2.91940-0.74002\imo$&$3.39289-0.84994\imo$&$3.86341-0.95235\imo$\\
$0.2$&$1.38107-0.33291\imo$&$1.87596-0.47514\imo$&$2.35219-0.60449\imo$&$2.81847-0.72326\imo$&$3.27906-0.83327\imo$&$3.73627-0.93591\imo$\\
     &$1.38107-0.33291\imo$&$1.87597-0.47513\imo$&$2.35221-0.60447\imo$&$2.81850-0.72323\imo$&$3.27911-0.83323\imo$&$3.73635-0.93586\imo$\\
$0.3$&$1.27723-0.30908\imo$&$1.74877-0.44702\imo$&$2.20360-0.57433\imo$&$2.64868-0.69214\imo$&$3.08771-0.80175\imo$&$3.52289-0.90433\imo$\\
     &$1.27724-0.30908\imo$&$1.74877-0.44701\imo$&$2.20361-0.57431\imo$&$2.64871-0.69212\imo$&$3.08776-0.80172\imo$&$3.52296-0.90429\imo$\\
$0.4$&$1.14122-0.27698\imo$&$1.57512-0.40612\imo$&$1.99630-0.52809\imo$&$2.40957-0.64273\imo$&$2.81743-0.75044\imo$&$3.22147-0.85188\imo$\\
     &$1.14122-0.27698\imo$&$1.57513-0.40611\imo$&$1.99631-0.52808\imo$&$2.40959-0.64271\imo$&$2.81747-0.75042\imo$&$3.22152-0.85185\imo$\\
$0.5$&$0.98025-0.23821\imo$&$1.36231-0.35313\imo$&$1.73596-0.46457\imo$&$2.10458-0.57170\imo$&$2.46965-0.67413\imo$&$2.83196-0.77182\imo$\\
     &$0.98025-0.23821\imo$&$1.36231-0.35313\imo$&$1.73597-0.46456\imo$&$2.10460-0.57169\imo$&$2.46967-0.67412\imo$&$2.83199-0.77179\imo$\\
$0.6$&$0.80110-0.19466\imo$&$1.11935-0.29056\imo$&$1.43247-0.38547\imo$&$1.74298-0.47889\imo$&$2.05196-0.57025\imo$&$2.35987-0.65912\imo$\\
     &$0.80110-0.19466\imo$&$1.11935-0.29055\imo$&$1.43248-0.38547\imo$&$1.74299-0.47888\imo$&$2.05198-0.57024\imo$&$2.35989-0.65910\imo$\\
$0.7$&$0.60947-0.14800\imo$&$0.85494-0.22164\imo$&$1.09744-0.29526\imo$&$1.33861-0.36881\imo$&$1.57929-0.44211\imo$&$1.81992-0.51489\imo$\\
     &$0.60946-0.14801\imo$&$0.85494-0.22164\imo$&$1.09744-0.29526\imo$&$1.33862-0.36881\imo$&$1.57930-0.44210\imo$&$1.81993-0.51488\imo$\\
$0.8$&$0.40990-0.09947\imo$&$0.57648-0.14915\imo$&$0.74150-0.19890\imo$&$0.90581-0.24880\imo$&$1.06987-0.29887\imo$&$1.23394-0.34909\imo$\\
     &$0.40990-0.09947\imo$&$0.57648-0.14915\imo$&$0.74150-0.19890\imo$&$0.90581-0.24880\imo$&$1.06987-0.29887\imo$&$1.23395-0.34909\imo$\\
$0.9$&$0.20588-0.04994\imo$&$0.28997-0.07491\imo$&$0.37338-0.09989\imo$&$0.45650-0.12490\imo$&$0.53948-0.14995\imo$&$0.62239-0.17505\imo$\\
     &$0.20588-0.04994\imo$&$0.28997-0.07491\imo$&$0.37338-0.09989\imo$&$0.45650-0.12490\imo$&$0.53948-0.14995\imo$&$0.62239-0.17505\imo$\\

\hline
\end{tabular}
\end{table*}

\begin{table*}
\caption{Fundamental modes ($l=2$, $Q=0$) of gravitational perturbations of vector($"-"$) type calculated by the
continued fraction method (the first value) and by the time-domain simulation (the second value).}\label{vector-RQNMs}
\begin{tabular}{|l|c|c|c|c|c|c|}
\hline
$\rho$&$D=5$&$D=6$&$D=7$&$D=8$&$D=9$&$D=10$\\
\hline
$0  $&$1.13400-0.32752\imo$&$1.52466-0.47412\imo$&$1.93446-0.61231\imo$&$2.35883-0.73874\imo$&$2.79281-0.85415\imo$&$3.23339-0.96031\imo$\\
     &$1.13400-0.32752\imo$&$1.52467-0.47412\imo$&$1.93447-0.61230\imo$&$2.35885-0.73871\imo$&$2.79285-0.85412\imo$&$3.23345-0.96025\imo$\\
$0.1$&$1.11435-0.32445\imo$&$1.49999-0.47120\imo$&$1.90445-0.60980\imo$&$2.32356-0.73670\imo$&$2.75235-0.85255\imo$&$3.18778-0.95907\imo$\\
     &$1.11435-0.32445\imo$&$1.50000-0.47119\imo$&$1.90447-0.60979\imo$&$2.32358-0.73668\imo$&$2.75238-0.85252\imo$&$3.18784-0.95902\imo$\\
$0.2$&$1.05738-0.31448\imo$&$1.42675-0.46085\imo$&$1.81478-0.60035\imo$&$2.21794-0.72854\imo$&$2.63114-0.84561\imo$&$3.05117-0.95321\imo$\\
     &$1.05738-0.31448\imo$&$1.42676-0.46084\imo$&$1.81479-0.60034\imo$&$2.21796-0.72852\imo$&$2.63117-0.84558\imo$&$3.05122-0.95316\imo$\\
$0.3$&$0.96883-0.29608\imo$&$1.30898-0.43923\imo$&$1.66798-0.57863\imo$&$2.04364-0.70827\imo$&$2.43048-0.82708\imo$&$2.82473-0.93628\imo$\\
     &$0.96883-0.29608\imo$&$1.30898-0.43923\imo$&$1.66798-0.57862\imo$&$2.04365-0.70826\imo$&$2.43051-0.82706\imo$&$2.51248-0.89812\imo$\\
$0.4$&$0.85670-0.26872\imo$&$1.15581-0.40298\imo$&$1.47213-0.53811\imo$&$1.80689-0.66732\imo$&$2.15518-0.78734\imo$&$2.51245-0.89815\imo$\\
     &$0.85671-0.26872\imo$&$1.15581-0.40297\imo$&$1.47213-0.53810\imo$&$1.80690-0.66731\imo$&$2.15520-0.78732\imo$&$2.12471-0.82648\imo$\\
$0.5$&$0.72870-0.23352\imo$&$0.97927-0.35226\imo$&$1.24192-0.47539\imo$&$1.52193-0.59819\imo$&$1.81754-0.71592\imo$&$2.12469-0.82650\imo$\\
     &$0.72870-0.23352\imo$&$0.97927-0.35225\imo$&$1.24192-0.47539\imo$&$1.52194-0.59818\imo$&$1.81755-0.71591\imo$&$1.68348-0.71083\imo$\\
$0.6$&$0.59065-0.19237\imo$&$0.78984-0.29049\imo$&$0.99434-0.39341\imo$&$1.21066-0.49989\imo$&$1.44075-0.60671\imo$&$1.68347-0.71085\imo$\\
     &$0.59065-0.19237\imo$&$0.78984-0.29049\imo$&$0.99434-0.39341\imo$&$1.21066-0.49988\imo$&$1.44075-0.60671\imo$&$2.82477-0.93624\imo$\\
$0.7$&$0.44658-0.14710\imo$&$0.59441-0.22174\imo$&$0.74231-0.29939\imo$&$0.89505-0.38063\imo$&$1.05542-0.46492\imo$&$1.22478-0.55081\imo$\\
     &$0.44658-0.14710\imo$&$0.59441-0.22174\imo$&$0.74231-0.29938\imo$&$0.89505-0.38063\imo$&$1.05543-0.46492\imo$&$1.22479-0.55081\imo$\\
$0.8$&$0.29911-0.09923\imo$&$0.39673-0.14920\imo$&$0.49222-0.20025\imo$&$0.58809-0.25289\imo$&$0.68581-0.30743\imo$&$0.78648-0.36396\imo$\\
     &$0.29911-0.09923\imo$&$0.39673-0.14920\imo$&$0.49222-0.20025\imo$&$0.58809-0.25289\imo$&$0.68581-0.30743\imo$&$0.78648-0.36396\imo$\\
$0.9$&$0.14990-0.04991\imo$&$0.19842-0.07491\imo$&$0.24523-0.10006\imo$&$0.29132-0.12542\imo$&$0.33718-0.15110\imo$&$0.38309-0.17716\imo$\\
     &$0.14990-0.04991\imo$&$0.19842-0.07491\imo$&$0.24523-0.10006\imo$&$0.29132-0.12542\imo$&$0.33718-0.15110\imo$&$0.38309-0.17716\imo$\\
\hline
\end{tabular}
\end{table*}

\begin{table*}
\caption{Fundamental modes ($l=2$, $\Lambda=0$) of gravitational perturbations of tensor type calculated by the
continued fraction method.}\label{tensorQQNMs}
\begin{tabular}{|l|c|c|c|c|c|c|}
\hline
$Q$&$D=5$&$D=6$&$D=7$&$D=8$&$D=9$&$D=10$\\
\hline
$0  $&$1.51057-0.35754\imo$&$2.01153-0.50194\imo$&$2.49678-0.63188\imo$&$2.97469-0.75056\imo$&$3.44882-0.86013\imo$&$3.92095-0.96212\imo$\\
$0.1$&$1.50687-0.35527\imo$&$2.00760-0.49926\imo$&$2.49270-0.62893\imo$&$2.97048-0.74743\imo$&$3.44451-0.85686\imo$&$3.91654-0.95876\imo$\\
$0.2$&$1.49575-0.34858\imo$&$1.99579-0.49136\imo$&$2.48041-0.62030\imo$&$2.95782-0.73829\imo$&$3.43153-0.84736\imo$&$3.90328-0.94898\imo$\\
$0.3$&$1.47711-0.33773\imo$&$1.97601-0.47872\imo$&$2.45985-0.60659\imo$&$2.93664-0.72389\imo$&$3.40980-0.83247\imo$&$3.88107-0.93374\imo$\\
$0.4$&$1.45084-0.32326\imo$&$1.94822-0.46216\imo$&$2.43101-0.58890\imo$&$2.90695-0.70552\imo$&$3.37938-0.81368\imo$&$3.85000-0.91466\imo$\\
$0.5$&$1.41684-0.30600\imo$&$1.91251-0.44296\imo$&$2.39413-0.56881\imo$&$2.86911-0.68503\imo$&$3.34072-0.79303\imo$&$3.81063-0.89397\imo$\\
$0.6$&$1.37526-0.28717\imo$&$1.86940-0.42278\imo$&$2.35002-0.54830\imo$&$2.82423-0.66461\imo$&$3.29520-0.77285\imo$&$3.76452-0.87411\imo$\\
$0.7$&$1.32676-0.26838\imo$&$1.82018-0.40345\imo$&$2.30048-0.52922\imo$&$2.77447-0.64603\imo$&$3.24525-0.75485\imo$&$3.71440-0.85670\imo$\\
$0.8$&$1.27290-0.25118\imo$&$1.76696-0.38619\imo$&$2.24787-0.51240\imo$&$2.72233-0.62977\imo$&$3.19345-0.73917\imo$&$3.66296-0.84114\imo$\\
$0.9$&$1.21589-0.23645\imo$&$1.71184-0.37127\imo$&$2.19317-0.49714\imo$&$2.66790-0.61582\imo$&$3.14034-0.72614\imo$&$3.59458-0.78869\imo$\\
\hline
\end{tabular}
\end{table*}

\begin{table*}
\caption{Fundamental modes ($l=2$, $Q=0$) of gravitational perturbations of tensor type calculated by the
continued fraction method.}\label{tensorRQNMs}
\begin{tabular}{|l|c|c|c|c|c|c|}
\hline
$\rho$&$D=5$&$D=6$&$D=7$&$D=8$&$D=9$&$D=10$\\
\hline
$0  $&$1.51057-0.35754\imo$&$2.01153-0.50194\imo$&$2.49678-0.63188\imo$&$2.97469-0.75056\imo$&$3.44882-0.86013\imo$&$3.92095-0.96212\imo$\\
$0.1$&$1.48334-0.35388\imo$&$1.98035-0.49836\imo$&$2.46129-0.62843\imo$&$2.93467-0.74725\imo$&$3.40415-0.85698\imo$&$3.87154-0.95914\imo$\\
$0.2$&$1.40454-0.34210\imo$&$1.88747-0.48630\imo$&$2.35494-0.61657\imo$&$2.81464-0.73573\imo$&$3.27015-0.84585\imo$&$3.72335-0.94843\imo$\\
$0.3$&$1.28232-0.32055\imo$&$1.73684-0.46249\imo$&$2.17943-0.59219\imo$&$2.61539-0.71141\imo$&$3.04733-0.82184\imo$&$3.47685-0.92487\imo$\\
$0.4$&$1.12818-0.28852\imo$&$1.53784-0.42371\imo$&$1.94114-0.55008\imo$&$2.34098-0.66777\imo$&$2.73842-0.77758\imo$&$3.13409-0.88045\imo$\\
$0.5$&$0.95396-0.24744\imo$&$1.30438-0.36922\imo$&$1.65303-0.48674\imo$&$2.00213-0.59880\imo$&$2.35173-0.70505\imo$&$2.70153-0.80566\imo$\\
$0.6$&$0.76899-0.20040\imo$&$1.05127-0.30195\imo$&$1.33321-0.40307\imo$&$1.61764-0.50244\imo$&$1.90496-0.59910\imo$&$2.19479-0.69251\imo$\\
$0.7$&$0.57901-0.15064\imo$&$0.79008-0.22746\imo$&$0.99988-0.30527\imo$&$1.21128-0.38368\imo$&$1.42542-0.46209\imo$&$1.64262-0.53993\imo$\\
$0.8$&$0.38680-0.10026\imo$&$0.52685-0.15100\imo$&$0.66493-0.20235\imo$&$0.80289-0.25437\imo$&$0.94164-0.30704\imo$&$1.08169-0.36030\imo$\\
$0.9$&$0.19359-0.05003\imo$&$0.26340-0.07513\imo$&$0.33185-0.10033\imo$&$0.39975-0.12564\imo$&$0.46744-0.15110\imo$&$0.53512-0.17672\imo$\\
\hline
\end{tabular}
\end{table*}

The existing literature on vector and tensor type of gravitational
perturbations is much fuller for the Schwarzschild case
\cite{Konoplya:2003dd,cardosoJHEP,Rostworowski:2006bp}, . In addition the
effective potential is positive definite for these types of
perturbations, so we do not suspect instability in them. The
fundamental modes for vector type $"+"$ gravitational
perturbations are given in Table \ref{vector+QQNMs}: the real
oscillation frequency of the dominating mode is growing as the
charge $Q$ grows, and after reaching some maximum, is decaying.
For $D=5$, the maximum of $Re \omega$ happens for $Q \approx 0.5
Q_{extr}$. This value of charge, at which $Re \omega$ has maximum,
is increasing for higher $D$, and is about $Q \approx 0.7
Q_{extr}$ for $D=10$. The imaginary part of the frequency is
monotonically decreasing as $Q$ is increasing for vector type
$"+"$ perturbations for all $D$.

The vector type $"-"$ of gravitational perturbations for
Reissner-Nordstr\"om case has rather different behavior (see Table
\ref{vector+QQNMs}): both $Re
\omega$ and $Im \omega$ are monotonically decaying as a function of $Q$.
Let us note that for vector type of gravitational perturbations we
perfectly reproduce the results of the pure Schwarzschild limit of
the previous papers \cite{Rostworowski:2006bp,cardosoJHEP,Konoplya:2003dd}. Thus, for example, for
$D=5$ we get in Table \ref{vector+QQNMs}: $\omega =
1.13400-0.32752\imo$ for vector type $"-"$ of gravitational
perturbations obtained with the time-domain approach, while the
Frobenius method in \cite{cardosoJHEP} give $\omega =
1.1340-0.3275\imo$. Let us remember that for vector type of
gravitational perturbations there are the two types $"-"$ and
$"+"$ of perturbations, because in fact we have the perturbations
of the Maxwell field and of gravitational field, which are
coupled. Thereby the type $"+"$ of vector perturbations
reduces to the perturbations of the test Maxwell field when the
electromagnetic background vanishes $Q=0$.

Dependence of the QNMs on the values of the $\Lambda$-term
parameter is the same for both types of vector perturbations: both
$Im \omega$ and $Re \omega$ are monotonically decreasing when
$\Lambda$ is increasing.

The tensor type of gravitational perturbations is known to
coincide with perturbations of test scalar field
\cite{Kodama:2003kk}, and the corresponding quasinormal spectrum
is isospectral as well \cite{Konoplya:2003dd}. The fundamental
modes for pure Reissner-Nordstr\"om and Schwarzschild-de Sitter
cases are given in Tables VII and VIII. Both real and imaginary
parts of quasinormal frequencies are monotonically decreasing as
charge or $\Lambda$-term is increasing.

This way one has quasinormal modes for all types of gravitational
perturbations analyzed for Reissner-Nordstr\"om-de Sitter black
holes.

\section{Conclusions}

Multidimensional black holes are subject of intensive research in
brane-world theories during recent years, because they are
expected to be produced in particles collisions in Large Hadron
Collider or in cosmic showers experiments. The basic feature one
needs to provide reliability of the black hole metric is
stability. Yet the stability of charged multidimensional black
holes with charge and $\Lambda$-term for $D \geq 8$ was an open
question until the present research. We have performed here an
extensive  time and frequency domains investigation of a response
of multidimensional black holes with arbitrary charge and
$\Lambda$-term and showed that the response is decaying and
thereby the black holes under consideration are stable. As a by
product we presented a detailed data on quasinormal frequencies
for the considered black holes and the dependence of the
frequencies on such parameters as black hole charge and the value
of the  cosmological term.

The are a few important points beyond our consideration: first of
all, it is Gauss-Bonnet generalizations of the Einstein solutions
\cite{Konoplya:2005sy,Gleiser:2005ra}, where an
instability can happen even in asymptotically flat case
\cite{Gleiser-new} in $D=5$ case. There numerical analysis of the gravitational perturbations of
the Gauss-Bonnet black holes for different $D$ is in progress
\cite{AKZ1}. Another important and interesting case is
asymptotically anti-de Sitter charged black holes, when the
instability also exists \cite{Gubser} and can be interpreted as a
phase transition in the dual temperature conformal field theory
\cite{KSZ}.

Apparently, an interesting alternative model for higher
dimensional black holes as a window in extra dimensions is the
Kaluza-Klein black holes with squashed horizons \cite{Jiro}. The
gravitational perturbations for these solution would be especially
actual, as they would probably show gravitational instability of
the Gregory-Laflamme type.

\begin{acknowledgments}
 This work was supported by \emph{Funda\c{c}\~{a}o de Amparo
\`{a} Pesquisa do Estado de S\~{a}o Paulo (FAPESP)}, Brazil.
\end{acknowledgments}

\end{document}